# Unstable Slip in Fault Gouge Driven by Temperature and Water

Li Wang[1], Jie Meng[2*], Dongpo Wang[3*], Gongji Zhang[2], Helge Hellevang[4]

[1]College of Architecture and Civil Engineering, Xi'an University of Science and Technology, Xi'an, China

[2]Faculty of Engineering, China University of Geosciences, Wuhan, China

[3]State Key Laboratory of Geohazard Prevention and Geoenvironment Protection, Chengdu University of Technology, Chengdu, Sichuan, China

[4]Department of Geosciences, University of Oslo, Oslo, Norway

*Corresponding author:

    Jie Meng (mengjie @cug.edu.cn)

    Dongpo Wang (wangdongpo@cdut.edu.cn)

**Abstract**

Microscale granular sliding within fault gouge is fundamental to earthquake nucleation, yet the mechanism by which temperature affects friction through interfacial water remains poorly understood. Here, large-scale molecular dynamics simulations were conducted on a hydrophilic quartz-water-quartz interface over 300-500 K to quantify temperature-dependent changes in frictional strength, real contact area, and water-layer structure. Results show that both the friction coefficient and friction force decrease monotonically with increasing temperature, following near-linear relationships of $\mu \propto T^{-1}$ and $F_t \propto A$, indicating that frictional weakening is primarily governed by temperature-driven contact restructuring. Structural analyses further show that heating progressively disrupts the hydrogen-bond network in the first adsorption layer, reduces adsorption-layer density, and weakens radial distribution peaks, demonstrating a transition of interfacial water from an ordered, strongly adsorbed state to a more diffuse, weakly bound configuration with delayering and quasi-phase-transition behavior. This interfacial reconstruction weakens intergranular bridging and structural cohesion, promoting a shift from structural locking to water-mediated lubrication. These results suggest that frictional stability under coupled temperature-water conditions is strongly controlled by the thermal evolution of interfacial water structure.

# 1 Introduction

The frictional behavior of deep faults and fault gouge is crucial for controlling earthquake nucleation and triggering, fault stability evolution, and the occurrence of geological engineering disasters (Marton et al., 2005; Scholz, 2019). Although the mechanical behaviors of slow slip in subduction zones, the periodic stick–slip of active faults, and transient instability under high-speed sliding are complex, their mechanical characteristics can be traced to the changes in frictional resistance at the fault interface and its sensitivity to external conditions (Lachenbruch, 1980; Ruina, 1983; Schwartz and Rokosky, 2007; Lee et al., 2024). These behaviors are influenced by multiple factors, with temperature increasingly recognized as a key environmental control on frictional and mechanical responses (Barbot, 2022; Boyet et al., 2023; Trowbridge et al., 2024). Current experimental studies have reported that temperature not only changes the frictional strength of fault materials but also may induce a transition from stable creep to unstable stick–slip (Dieterich, 1978; Yao et al., 2023; Ashman et al., 2025), especially in deep earthquakes, deep underground engineering, and high-speed sliding thermal softening phenomena (Tse and Rice, 1986; Goebel et al., 2024; Yan et al., 2025).

The friction coefficient response of fault materials to temperature clearly exhibits stage characteristics. Most fault materials exhibit high and relatively stable friction coefficients under low-temperature or dry conditions; these findings are consistent with the predictions of Byerlee's law (Brace and Byerlee, 1966; Stesky et al., 1974; Aharonov and Scholz, 2018). As the temperature increases to a specific range, the frictional strength generally decreases, which may be accompanied by a transition from stable creep to unstable stick–slip (Blanpied et al., 1995; He et al., 2006; Mitchell et al., 2015). More importantly, fault slip is closely related to pore fluid (Jamison and Teufel, 1979; Teufel, 1981; Lockner and Byerlee, 1994; Yuan et al., 2024), and fault gouge with high water contents is more sensitive to temperature (Andrews, 2002; Wibberley and Shimamoto, 2005; Ferri et al., 2010; Sawai et al., 2016; Badt et al., 2020). The thermal pressurization mechanism proposed by Andrews (2002) and Wibberley and Shimamoto (2005) suggests that the heating expansion of pore water can produce transient high pressures during rapid sliding, significantly reducing frictional resistance. Thermal–hydro shear experiments on natural fault gouge indicate that increasing temperature can decrease the cohesion of material and increase lubrication during sliding. This can result in velocity-weakening behavior (Boulton et al., 2014; Moore et al., 2016; Boulton et al., 2018).Similar trends have been observed for the Alps, South China Sea Trench, and San Andreas fault gouge (den Hartog et al., 2012; Tesei et al., 2014; Moore et al., 2016; Niemeijer et al., 2016). Although both field and laboratory studies have reported that fault frictional behavior is extremely sensitive to thermal–hydro conditions, the microscopic physical basis for its temperature dependence has not been completely analyzed (Mei and Rudnicki, 2023; Nie and Barbot, 2024).

The competitive relationship between the transition and microscopic mechanisms of friction behavior, especially the thermal dependence between unstable sliding processes, is very similar (Chester, 1994; Chen and Spiers, 2016; Tian et al., 2018).

Mei and Rudnicki (2023) reported that the sensitivity of microphysical parameters to temperature determines the steady-state friction curve and its stability boundary through a wet granite fault gouge model. Okuda et al. (2023) conducted thermal–hydro friction experiments and microslice structural analysis. The test results revealed that pore water promotes the formation of small sodium feldspar and chlorite particles under thermal conditions. Pore water can clearly reduce the friction strength in shear zones, which results in increasing fault instability at the macroscopic scale. Guvercin et al. (2025) reported that the friction stability of POR shale is closely related to the thermally activated transition process under lower crust conditions and that the activation transition temperature is controlled by the frictional structural characteristics of the particles. Notably, an increasing number of observations, experiments, and simulations indicate that the transition from stable to unstable sliding stems from the competition between multiple microscopic deformation mechanisms of fault materials (Eijsink et al., 2022; Nie and Barbot, 2024; Barbot et al., 2025; Venegas-Aravena, 2025; Wu and Barbot, 2025). Sliding responses often begin at the initial contacts between grains, with recent hypotheses suggesting that temperature directly alters the molecular state and structure of interfacial water (Chowdhury and Ghasemi-Fare, 2024; Ashman et al., 2025), affecting both contact properties and frictional behavior. Notably, the role of water in friction is highly complex; it does not always act as a lubricant (Chen and Qian, 2021; Gao et al., 2025). Under humid conditions, the frictional resistance can even reach twice that observed under dry conditions (Hasz et al., 2018; Peng et al., 2022), indicating a distinct strengthening effect. Furthermore, the coupling between temperature and water content renders the underlying frictional mechanisms even more complex and less predictable. Currently, our understanding of how temperature–water variations regulate frictional behavior is primarily derived from macroscopic experimental correlations (Okamoto et al., 2019; Belzer and French, 2022, 2024; Ashman et al., 2025), which lack direct mechanistic data at the grain scale. Consequently, the microscale evolution of the interfacial water structure and its regulatory role in frictional instability remain poorly understood.

To overcome the limitations of macroscopic experiments in resolving the evolution of interfacial water structure and its transient slip response at the grain scale, a representative hydrophilic grain–grain contact system is constructed, and molecular dynamics simulations are employed to systematically investigate how temperature regulates the coupled evolution of interfacial water structure and frictional response. Compared with conventional laboratory approaches, molecular simulations can directly capture the dynamic reorganization of interfacial structures at the atomic scale, providing an effective method for elucidating the microscopic origins of a wide range of macroscopic frictional behaviors (Sakuma et al., 2018; He et al., 2022; Bui et al., 2023). Under well-controlled normal loading and sliding conditions, key variables—including the transient frictional response, contact evolution, and the hydrogen bond network—are quantitatively characterized across a range of temperatures. This study elucidates how heating drives the interface to transition from a structurally constrained (locked) state to a water-mediated lubricated state, thereby providing mechanistic insight into unstable slip in faults and landslides.

**2 Methods**

2.1 Molecular model for interfacial sliding and the force field

In deep fault zones and cataclastic zones, quartz ($SiO_2$) is a major mineral in most brittle fault rocks as well as in sandstones and granites (Blenkinsop and Drury, 1988; Giger et al., 2007; Masuda et al., 2019; Scuderi et al., 2025). Its interfacial structure and frictional behavior are therefore expected to exert first-order control on slip stability. A symmetric quartz–water–quartz trilayer system was thus constructed (Fig. 1) to represent the contact and sliding response of quartz grains under coupled temperature–water conditions. Figure 1a shows a schematic of the displacement–time response and a typical fault–shear architecture, highlighting that heating-induced instability may originate from abrupt changes in the state of interfacial water. Figure 1b presents the atomistic configuration of the quartz–water–quartz contact. The model consists of two α-quartz crystals exposing the (001) surface with a confined water layer in between, forming a representative hydrophilic interface.

The lower quartz block is divided, from bottom to top, into a rigid layer, a thermostatted layer, and a Newtonian layer. The upper quartz asperity is represented by a rigid spherical cap with a radius of 36 Å and a cap height of 30 Å; the top 915 atoms of this asperity are defined as the loading layer. Atoms in the Newtonian layer evolve under classical dynamics and constitute the primary region for interfacial frictional response and water-structure reorganization. The thermostatted layer controls the system temperature, whereas the rigid layer serves as a fixed substrate to prevent unintended motion of the lower block. The loading layer is used to apply a normal load and to impose velocity-driven shear through a virtual spring, from which the shear-force response is obtained.

Each friction simulation includes three stages. First, the system is fully relaxed in the NVT ensemble at 300 K for 1000 ps. Second, the temperature is increased to the target value (300, 350, 400, 450, and 500 K) using a Langevin thermostat over 200 ps. A constant normal load is then applied along the z direction to the loading layer of the upper asperity to establish contact, and the system is allowed to reach an initial mechanical steady state within an additional 200 ps at the target temperature. Third, sliding is initiated by applying a spring force along the x direction with a spring constant of 200 N/m. The driving velocity of the spring is set to 0.1 Å/ps. For each target temperature, ten stepwise normal loads ranging from 0 to 40 nN are imposed. The mean friction force under each load is calculated from simulations spanning at least 100 Å of sliding displacement, yielding the load-dependent friction curves at each temperature.

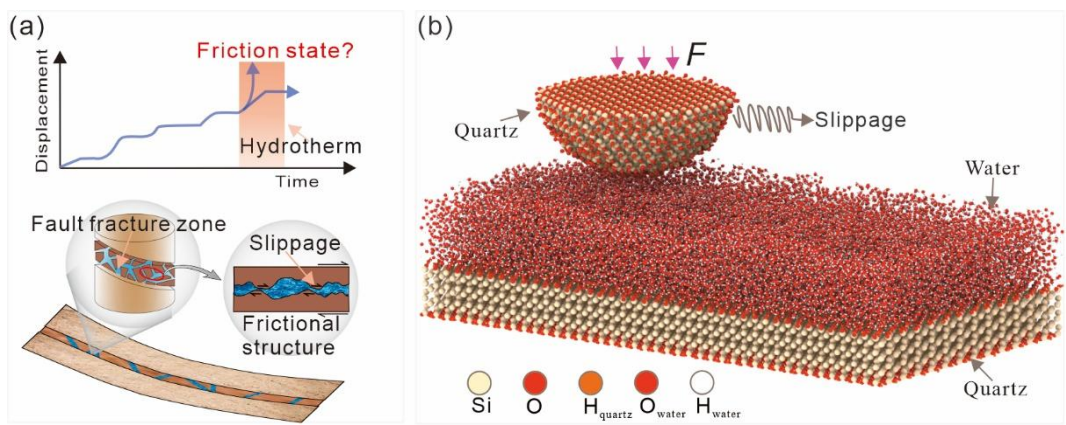

**Fig. 1.** Simulation framework and atomistic configuration for grain-scale friction

under coupled temperature–water conditions. (a) Schematic of thermally activated grain-scale frictional instability and microscale contacts, adapted from Nie and Barbot (2024). (b) Atomistic configuration of the quartz–water–quartz interfacial model.

2.2 Calculation of the friction force and real contact area

The interfacial friction force was obtained from the tensile force of a virtual spring attached to the rigid layer of the upper quartz block. When the driving velocity of the spring end matches the actual sliding velocity of the rigid layer, the spring force equals the true interfacial friction force. At the macroscopic scale, Amontons' law states that friction is proportional to the normal load. At the nanoscale, however, interfacial friction is governed by the real contact area and local bonding states. Consequently, the temperature dependence of friction reflects not only a shift between lubrication and adhesion but also the thermal sensitivity of interfacial water structure.

According to classical Amontons' laws, the macroscopic friction force is proportional to the applied load:

$$F_t = \mu L \tag{1}$$

where $F_t$ is the friction force, $\mu$ is the friction coefficient, and $L$ is the normal load. This formulation implies that $\mu$ is independent of the apparent contact area. At the microscopic scale, because frictional interfaces are not perfectly smooth, the apparent contact area consists of multiple discrete asperity contacts ($\sum A_{\text{asp}}$), and the friction force can be expressed as:

$$F_t = \tau \sum A_{\text{asp}} \tag{2}$$

where $\tau$ is the effective interfacial shear strength.

Real contact regions were identified using an interatomic-distance criterion. A contact was defined when the separation between opposing quartz atoms was smaller than a cutoff radius $R_{cut}$. The interface was discretised into equal-area grids, and the areas of all grids satisfying this criterion were summed to obtain the distance-based contact area $A_{dist}$. To determine an appropriate $R_{cut}$, $A_{dist}$ calculated under different cutoff values was compared with the contact area predicted by the Hertz model under the same normal load (Mo et al., 2009; Wang et al., 2022):

$$A_{Hertz} = \pi \left(\frac{3R}{4E^*}\right)^{2/3} F_N^{2/3} \qquad (3)$$

where $R$ is the sphere radius, $F_N$ is the applied normal load, and $E^*$ is the effective elastic modulus defined as:

$$E^* = \left(\frac{1-\nu_1^2}{E_1} + \frac{1-\nu_2^2}{E_2}\right)^{-1} \qquad (4)$$

Here, $E_1$ and $E_2$ are the Young's moduli of the two quartz bodies in contact, and $\nu_1$ and $\nu_2$ are their Poisson's ratios. This comparison ensures that the selected $R_{cut}$ yields an atomistic contact criterion consistent with the load-bearing capacity of the interface.

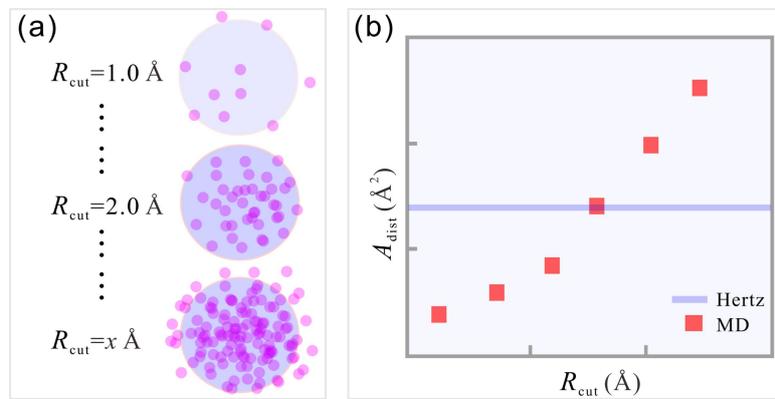

**Fig. 2.** Identification of interfacial contact area based on the interatomic-distance criterion. (a) Schematic illustration of atomistic contact regions identified under different cutoff radii $R_{cut}$. As $R_{cut}$ increases, the area recognized as contacting progressively expands. (b) Variation of the contact area $A_{dist}$ obtained from molecular simulations as a function of $R_{cut}$, compared with the Hertz-model prediction, which is used to determine the most appropriate cutoff criterion.

2.3 Structural parameters of the frictional interface

To further resolve the microscopic structural response of interfacial water during sliding, we quantified the hydrogen-bond network, interfacial water density, and radial distribution functions as key descriptors. These metrics capture the evolution of molecular connectivity and local ordering/adsorption stability, and are used to interpret the structural origins of temperature-driven transitions in frictional regimes.

**(1) Hydrogen-bond network**

Hydrogen bonds (Hbs) constitute the basic structural and energetic units of water and are essential for maintaining the cohesion and connectivity of interfacial water layer (Luzar and Chandler, 1996a). Under confined hydrophilic contacts, the integrity of the Hb network governs the ability of water to sustain molecular bridging and a structurally constrained adsorption layer, thereby modulating the effective shear resistance during sliding. As illustrated in Fig. 3, the H atom bonded to the donor oxygen forms a connection with the lone pair of electrons on the acceptor oxygen. H atoms with positive charges are attracted to lone pairs of electrons with negative charges. The geometric criteria for calculation of the water hydrogen bond were $r_{OO}$ < 3.5 Å and $\phi_{HOO}$ < 30° (Luzar and Chandler, 1993, 1996b). $r_{OO}$ is the O−O distance, and ϕ$_{HOO}$ is the angle between the O-H bond and $r_{OO}$ within the molecule.

**(2) Radial distribution function (RDF)**

To characterise the spatial organisation and short-range ordering of water near the interface and to evaluate their potential influence on interfacial adhesion and shear response, RDFs $g(r)$ were calculated for selected atom pairs from the molecular dynamics trajectories. The RDF is defined as the normalised probability density of finding a target atom within the shell $r$ to $r + dr$ around a reference atom relative to a uniform bulk distribution (Fig. 3):

$$g(r) = \frac{1}{4\pi r^2 \rho} \langle \sum_{i \neq j} \delta(r - r_{ij}) \rangle \tag{5}$$

where $\rho$ is the average number density and $r_{ij}$ is the instantaneous distance between atoms $i$ and $j$. To minimise bulk-water contributions, the RDF calculations were restricted to water molecules within 1.0 nm normal to the solid surface, and key pairs such as $O_w$–surface and $O_w$–$H_w$ were analysed. The position and height of the first peak reflect the coordination environment, local confinement, and adsorption strength of the interfacial layer, which together indicate the structural resistance of the water film to shear perturbation.

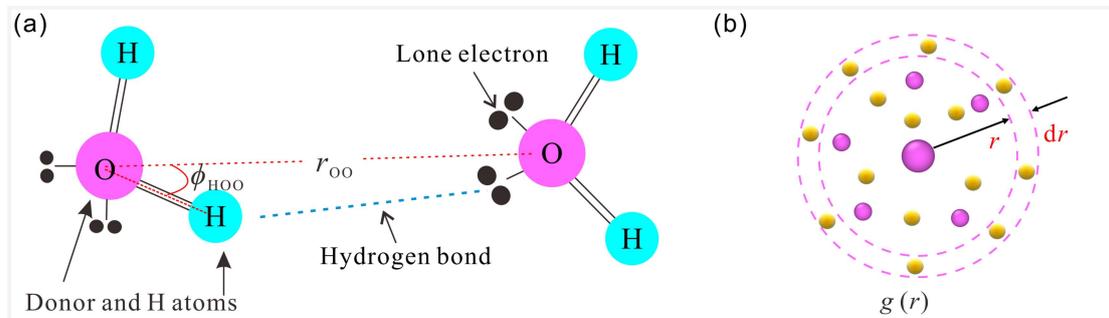

**Fig. 3.** Geometric definition of hydrogen bonding between water molecules. (a)

Hydrogen-bond geometry defined by the O–O distance $r_{OO}$ and the angle $\phi_{HOO}$, (b) Schematic of the RDF $g(r)$, showing the probability of finding neighboring molecules within a shell of thickness $dr$ at distance $r$.

**3. Results and Discussion**

3.1 Frictional behavior under different temperature conditions

To elucidate the overall effect of temperature on interfacial frictional behavior, the variations in the friction force, friction coefficient (μ), and real contact area were systematically examined over the temperature range of 300–500 K. As shown in Fig. 4a, the friction force increases approximately linearly with the normal load at all the examined temperatures, indicating that the nanoscale frictional response still conforms to the Amontons' (Barber, 2013) law–type load dependence observed at the macroscopic scale. Under a constant normal load, the friction force decreases systematically with increasing temperature, indicating a thermally induced reduction in interfacial shear resistance. This observation is consistent with previous experimental findings that interfacial water exerts a drag-enhancing effect (Hasz et al., 2018; Peng et al., 2022), whereas the present results further indicate that increased temperature substantially reduces this water-induced frictional enhancement. The temperature dependence of $\mu$ is quantitatively characterized in Fig. 4b. At room temperature, μ is comparable to the experimentally measured value; as the temperature increases, $\mu$ decreases monotonically and is clearly linearly correlated with the reciprocal of the temperature ($T^{-1}$). This $\mu$–$T^{-1}$ scaling relationship indicates that the thermal weakening of interfacial friction is controlled by a thermally activated process, in which thermal fluctuations gradually decrease the potential energy corrugation at the quartz–water interface, thereby reducing the shear strength that can be sustained per unit of real contact area.

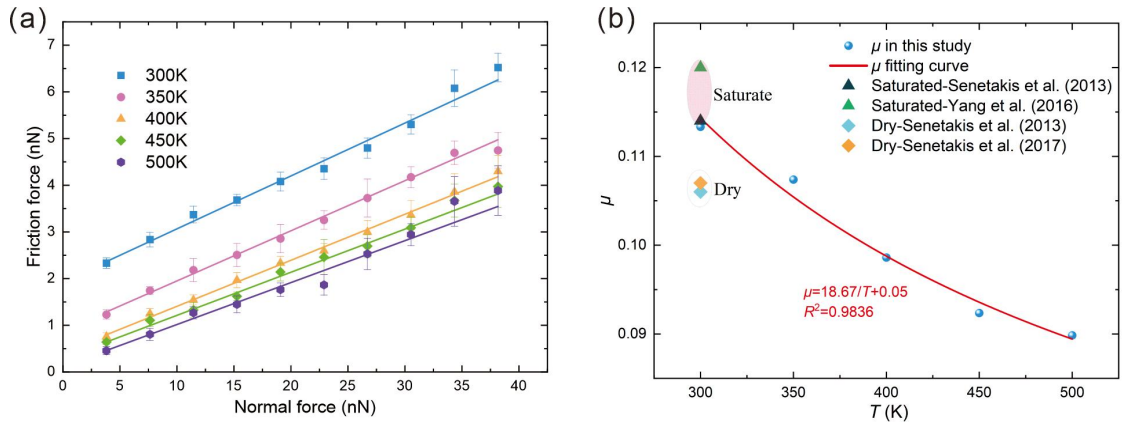

**Fig. 4** Load-dependent friction behavior and temperature dependence of $\mu$. (a) Friction force as a function of the normal load at different temperatures; (b) variation in $\mu$ with temperature, with experimental data at room temperature from Senetakis et al. (2013); Senetakis et al. (2017).

The evolution of the real contact area during sliding provides further data of the coupling between the interfacial structure and mechanical response. As shown in Fig. 5, the friction force exhibits quasiperiodic oscillations at all temperatures, accompanied by synchronous expansion and contraction of the real contact area. When the friction force increases toward a local peak, the interface is in an adhesion-dominated stage: the shear displacement is suppressed, the contact asperities and adsorbed water are compacted, and the local load-bearing capacity increases, resulting in contact area enlargement. When the shear stress exceeds the interfacial adhesion strength, adhesion junction rupture and sliding initiate. During this slip phase, local structural relaxation and contact breakdown cause a rapid reduction in the real contact area, corresponding to the simultaneous decrease in friction force to a valley.

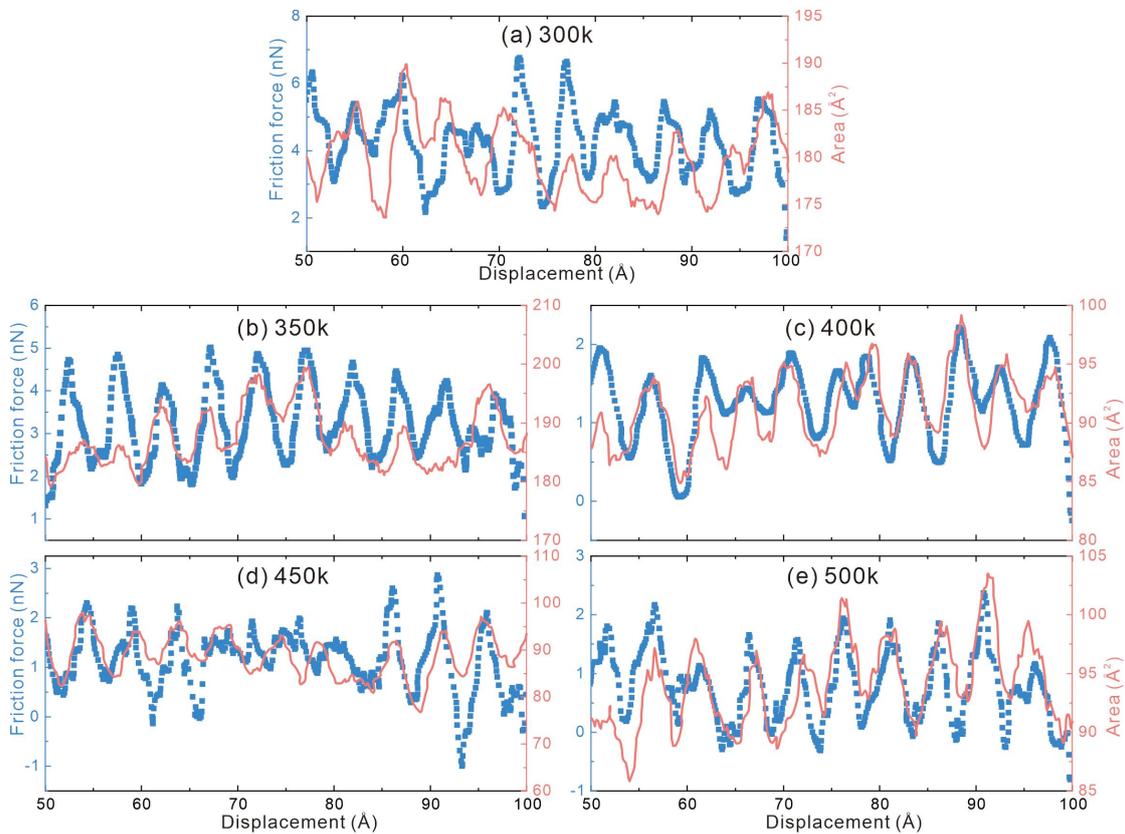

**Fig. 5** Simultaneous evolution of the friction force and real contact area during sliding at different temperatures.

The statistical correlation between the friction force and contact area is summarized in Fig. 6. Across all the temperatures, the friction force increases nearly linearly with the real contact area, revealing a strong positive correlation and indicating that the interfacial shear resistance is controlled primarily by the extent of real contact. The slope is greatest at 300 K and decreases gradually with increasing temperature. Thus, even at comparable contact areas, friction forces are significantly lower at higher temperatures. This trend indicates that elevated temperature reduces both the amplitude of contact area variations and the effective shear resistance that can be sustained per unit contact area.

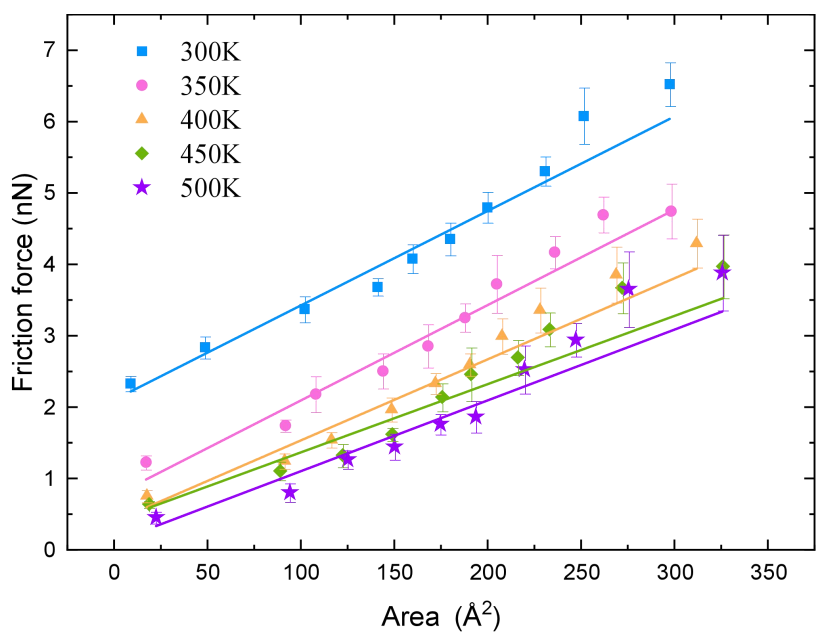

**Fig. 6** Correlations between the friction force and real contact area at different temperatures

3.2 Evolution of the frictional interface structure with increasing temperature

Building on the mechanical results that indicate a systematic reduction in interfacial shear resistance with increasing temperature, atomistic visualizations were performed to elucidate the structural origins of frictional weakening (Fig. 7). The top-view configurations reveal a distinct temperature-dependent expansion of the friction-induced disturbed zone. At 300 K, structural perturbations are confined to a narrow annulus along the contact perimeter. As the temperature increases to 350–400 K, the disturbed region spreads outward and transitions from a well-defined boundary to a more diffuse, spatially extended zone. At 450–500 K, large high-disturbance areas develop around the interface, indicating substantially larger atomic displacements and a greater spatial extent of friction-induced deformation. The front-view cross sections further reveal that at 300–350 K, the water layers above the interface remain densely packed with only minimal perturbations. Above 400 K, the disturbances propagate upward into deeper layers, and at 450–500 K, the water film exhibits localized swelling, structural loosening, and outward-scattering clusters, reflecting a distinct loss of interfacial structural stability.

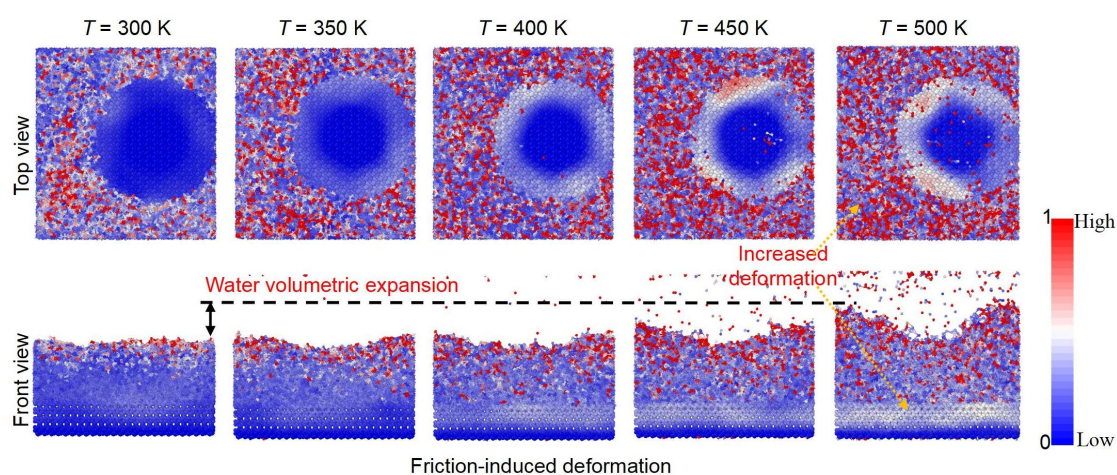

**Fig. 7** Temperature-dependent, friction-induced deformation and interfacial structural response. Top view: Atomic configurations of the contact region at different temperatures, showing the gradually increasing disturbed zone with increasing temperature. Front view: Cross-sectional views of the interface showing friction-induced deformation and the temperature-dependent expansion of the disturbed water layer.

Additional insight into the evolution of the interfacial water network is provided in Fig. 8. The three-dimensional hydrogen bond morphology (Fig. 8a) shows a nearly continuous and compact H-bond shell at 300 K, where the low–H-bond region (blue) is minimal, indicating a dense and robust hydrogen bond network. At 400 K, the

low–H-bond region increases, and the network becomes spatially irregular, indicating a reduction in both structural cohesion and resistance to shear perturbation. At 500 K, this region increases significantly, and the central interfacial zone collapses into a loose molecular configuration, revealing the breakdown of the hydrogen bond framework that ensures low-temperature stability. The quantitative statistics (Fig. 8b) reveal an approximately linear decrease in the number of hydrogen bonds from 300–500 K, confirming that the molecular connectivity and cohesion of the interfacial water layer gradually decrease with increasing temperature. These observations indicate that thermal weakening of the hydrogen bond network, degradation of layering, and localized collapse of the interfacial water film collectively reduce its structural support, adhesion, and resistance to shear-induced rearrangement. This structural destabilization closely parallels the temperature dependence of the friction force and $\mu$, providing direct microscopic data on thermally induced frictional weakening at hydrophilic mineral–water interfaces.

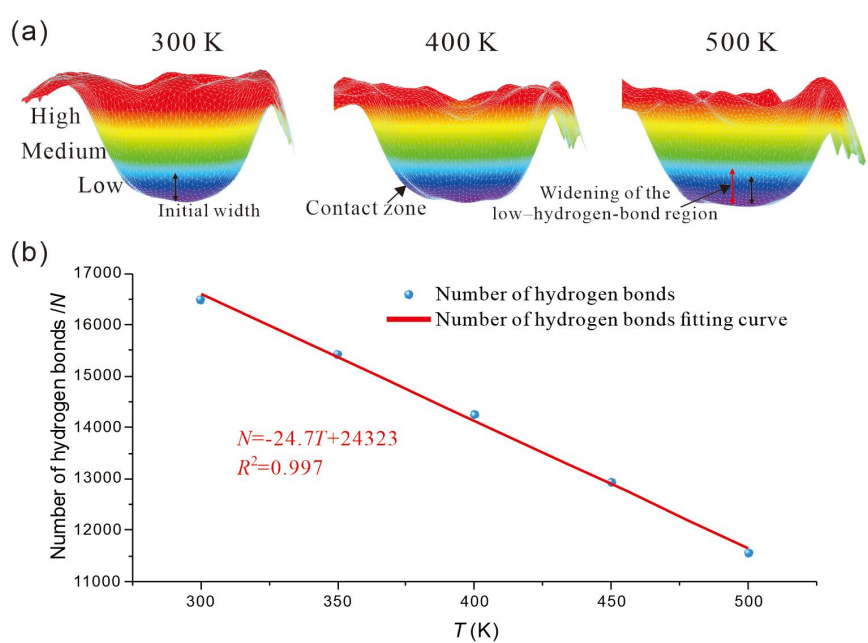

**Fig. 8** Hydrogen-bonding structure around the sliding body and its temperature-dependent evolution. (a) Three-dimensional representations of the hydrogen bond distribution around the slider at different temperatures; (b) variation in the total number of Hbs as a function of temperature, which monotonically decreases with increasing temperature.

3.3 Interfacial water evolution with increasing temperature

To elucidate how temperature controls the structure of interfacial water and its frictional consequences, two-dimensional projections of water molecules near the quartz–water interface at 300–500 K are presented in Fig. 9. At 300 K, water molecules exhibit a highly ordered, lattice-like arrangement. This "structural-locking" state is characterized by strong adhesion and high interfacial stiffness, which is consistent with elevated frictional strength and distinct stick–slip. At 400 K, the spatial uniformity of the adsorbed layer is noticeably reduced, and local loosening and depletion zones indicate a transition to a "quasifluidized" state. The weakened structural confinement promotes easier slip initiation and reduces the amplitude of frictional oscillations. By 500 K, the adsorbed water no longer retains a layered morphology; short-range order vanishes, and the molecules exhibit a quasidiffusive distribution. This "lubricating-medium" state substantially decreases solid–liquid coupling and the overall frictional resistance.

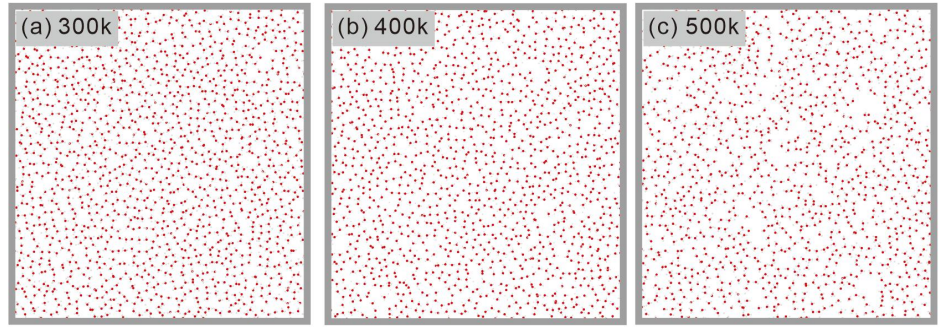

**Fig. 9** Near-surface water distribution characteristics at different temperatures

The corresponding density profiles (Fig. 10) further reveal temperature-driven changes to the interfacial water film. With increasing temperature, the molecular packing density near the solid surface decreases systematically, particularly within the first adsorption layer (2–4 Å), whose peak height decreases significantly. At 300–350 K, a dense and well-defined adsorbed layer is present, which is consistent with a stable hydrogen bond network and well-ordered molecular arrangement. This configuration provides strong interfacial support under shear, producing high friction levels and distinct serrated stick–slip signals. As the temperature increases to 400–500 K, the primary density peak decreases and widens, reflecting increased thermal motion, decreased adsorption, and relaxation of orientation constraints. The second and third layers (4–10 Å) progressively lose their oscillatory structure, marking the transition from layered coordination to bulk-like disorder. This thermally induced delayering directly results in mechanical behavior as a systematic reduction in frictional strength and smaller stick–slip amplitudes. As ordered water transforms into a looser, more mobile structure, lubrication is increased, molecular bridging decreases, and adhesion forces decrease, reducing energy dissipation and force fluctuations during sliding.

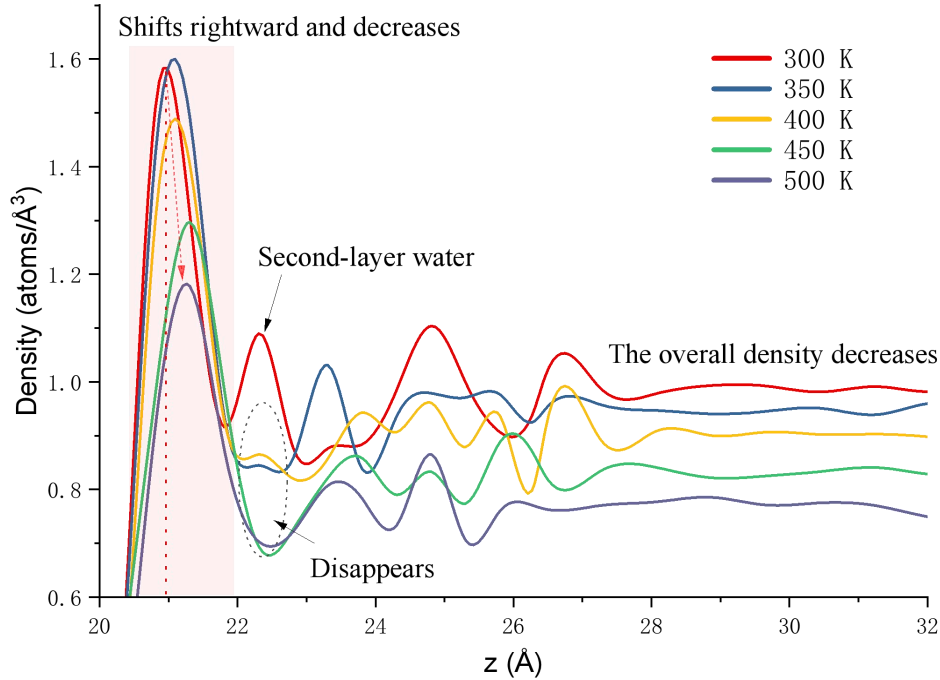

**Fig. 10** Temperature-dependent layering characteristics of interfacial water revealed by density profiles.

To further resolve the interfacial structural mechanisms responsible for frictional weakening, Fig. 11 shows the temperature evolution of RDFs for water–water pairs (Ow–Ow and Hw–Ow) and water–quartz pairs (Quartz–Ow and Quartz–Hw). With respect to Ow–Ow and Hw–Ow (Figs. 11a–b), the gradual decrease in the first-peak intensity with increasing temperature indicates a steady decrease in the short-range order within the water layer. Stronger thermal agitation increases intermolecular distance fluctuations and disrupts the hydrogen bond network (as also shown in Fig. 8), inducing a transition from a highly ordered "structural-locking" state to a more diffusive, lubrication-dominated configuration. This structural relaxation explains the reduction in the frictional peak frequency and amplitude and the increased continuity of sliding. Temperature also controls the interactions between water and the quartz surface. The Oq–Ow RDF peak increases with temperature, indicating that oxygen in water molecules more readily approaches surface hydroxyls—primarily via physical adsorption rather than stable bonding. Conversely, the Oq–Hw peak decreases monotonically, reflecting the progressive breakdown of Hbs and a reduction in surface adhesion. Together, these trends show that increasing temperature causes water to transition from an "ordered bridging" configuration to a "weakly adsorbed, diffusive" state, decreasing interfacial cohesion and both the friction force and $\mu$.

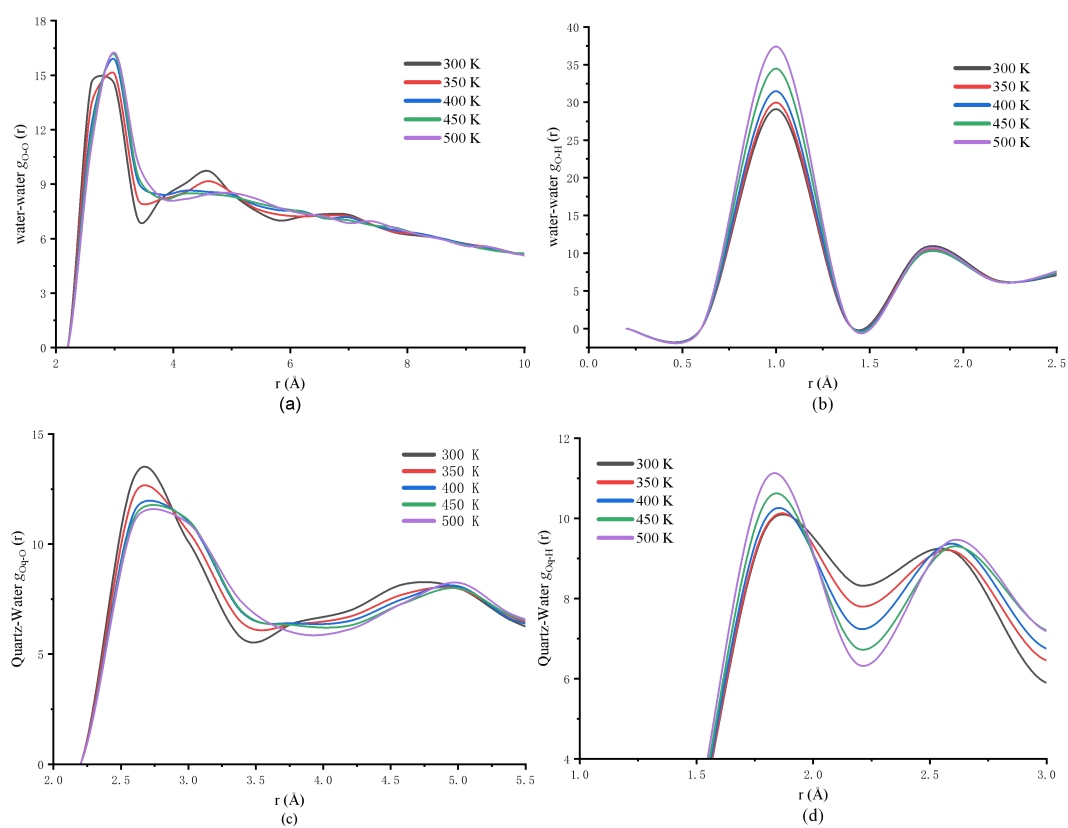

**Fig. 11** Temperature-dependent RDFs of interfacial water and water–quartz interactions.

Long-standing interpretations of temperature-induced fault weakening and increased slip instability have primarily emphasized grain-intrinsic processes, including mineral phase transitions, thermally activated grain softening, and thermal damage within shear zones (Allen, 1979; Han et al., 2007; Barbery et al., 2023). More recently, fault-slip studies have proposed an alternative perspective in which heating may not only alter the mechanical properties of grains but also affect slip stability through changes in the interfacial water state and water-structure evolution (Chowdhury and Ghasemi-Fare, 2024; Ashman et al., 2025). However, owing to limitations in experimental resolution and the intrinsic difficulty of capturing transient interfacial processes, direct data that links mechanical responses to microscopic structural evolution is lacking.

The current molecular-scale simulations address this limitation by resolving the temperature-dependent evolution of interfacial water, which remains challenging to assess experimentally, providing direct data for a microscopic connection between temperature and frictional instability. Before distinct weakening, interfacial water has a dense, layered adsorption structure that results in a high-stiffness, shear-constrained "structural locking" state, corresponding to elevated $\mu$ and distinct stick–slip signatures. With increasing temperature, the hydrogen bond integrity gradually decreases, and the adsorption strength decreases; the density peak of the first adsorption layer decreases, and delayering increases, whereas the interfacial water film transitions from an ordered adsorbed configuration to a more diffusive state, accompanied by the relaxation of interfacial constraints. This structural relaxation directly reduces the effective shear resistance provided per unit real contact area, for which $\mu$ serves as a parameterized measure (Estrada et al., 2008; Białas et al., 2021; He et al., 2021). Moreover, the observed linear $\mu$–$T^{-1}$ scaling indicates that slip initiation becomes increasingly controlled by thermally activated bond breaking and water layer reorganization events, facilitating a transition from a low-temperature adhesion-locked regime to a high-temperature thermally destabilized, quasi–velocity-weakening sliding mode.

The observed temperature-driven delayering and hydrogen bond degradation of interfacial water are consistent with the reductions in friction force, the decrease in stick–slip amplitude, and the temperature dependence of contact behavior. These results indicate that the effect of temperature on frictional stability cannot be attributed solely to grain-intrinsic thermal softening. Alongside mineral phase transitions and grain softening, this study provides molecular-scale evidence that the thermodynamic evolution of interfacial water can serve as a primary pathway by which heating promotes unstable slip. This implication is particularly relevant for deep faults, where cataclastic zones commonly contain hydrophilic minerals (Niwa et al., 2016; Dang and Zhou, 2021; Ma et al., 2024). Layered adsorbed water is thus expected to be widespread at grain contacts, and its temperature-induced structural loosening and reduced adhesion are likely to reduce slip thresholds and increase lubrication.

## 4. Conclusions

In this study, molecular dynamics simulations of a quartz–water–quartz hydrophilic interface system ranging from 300–500 K were conducted, and the effect of the dominant microscopic mechanism on frictional weakening in under coupled temperature–water conditions was revealed. The main conclusions reached are as follows: An increasing temperature systematically reduces interfacial friction, and the linear scaling of $\mu$–$T^{-1}$ indicates an increasingly thermally activated slip process. Across all temperatures, the friction force remains positively correlated with the real contact area. Although the contact area under high-temperature conditions is comparable, the friction forces are lower. This finding indicates that the interfacial mechanical response becomes increasingly affected by the evolving structural state of the interface rather than solely by geometric contact.

Temperature is a non-negligible concomitant variable during frictional sliding, evolving through the combined effects of frictional dissipation and heat transfer. Accordingly, the frictional stability of fault and landslide systems is governed not only by the thermal properties of the constituent minerals, but also—critically—by the thermodynamic evolution of interfacial water on hydrophilic surfaces. The molecular-scale data presented here provide a microphysical basis for understanding low friction and potentially unstable slip.

**Acknowledgements**

The research is supported by the National Natural Science Foundation of China (42472357), the Natural Science Foundation of Hubei Province of China (2025AFB380), the China Postdoctoral Science Foundation-Hubei Joint Support Program (2025T046HB), and the China Postdoctoral Science Foundation (2025M780439).

**Declaration of competing interest**

The authors declare that they have no known competing financial interests or personal relationships that could have appeared to influence the work reported in this paper.